\documentstyle[12pt,epsf]{article}

\def\hybrid{\topmargin -20pt    \oddsidemargin 0pt
        \headheight 0pt \headsep 0pt
        \textwidth 6.25in       
        \textheight 9.5in       
        \marginparwidth .875in
        \parskip 5pt plus 1pt   \jot = 1.5ex}

\hybrid

\def\baselinestretch{1.2}

\catcode`\@=11

\def\marginnote#1{}
\newcount\hour
\newcount\minute
\newtoks\amorpm
\hour=\time\divide\hour by60
\minute=\time{\multiply\hour by60 \global\advance\minute by-\hour}
\edef\standardtime{{\ifnum\hour<12 \global\amorpm={am}%
        \else\global\amorpm={pm}\advance\hour by-12 \fi
        \ifnum\hour=0 \hour=12 \fi
        \number\hour:\ifnum\minute<10 0\fi\number\minute\the\amorpm}}
\edef\militarytime{\number\hour:\ifnum\minute<10 0\fi\number\minute}

\def\draftlabel#1{{\@bsphack\if@filesw {\let\thepage\relax
   \xdef\@gtempa{\write\@auxout{\string
      \newlabel{#1}{{\@currentlabel}{\thepage}}}}}\@gtempa
   \if@nobreak \ifvmode\nobreak\fi\fi\fi\@esphack}
        \gdef\@eqnlabel{#1}}
\def\@eqnlabel{}
\def\@vacuum{}
\def\draftmarginnote#1{\marginpar{\raggedright\scriptsize\tt#1}}

\def\draft{\oddsidemargin -.5truein
        \def\@oddfoot{\sl preliminary draft \hfil
        \rm\thepage\hfil\sl\today\quad\militarytime}
        \let\@evenfoot\@oddfoot \overfullrule 3pt
        \let\label=\draftlabel
        \let\marginnote=\draftmarginnote
   \def\@eqnnum{(\theequation)\rlap{\kern\marginparsep\tt\@eqnlabel}%
\global\let\@eqnlabel\@vacuum}  }

\def\preprint{\twocolumn\sloppy\flushbottom\parindent 2em
        \leftmargini 2em\leftmarginv .5em\leftmarginvi .5em
        \oddsidemargin -.5in    \evensidemargin -.5in
        \columnsep .4in \footheight 0pt
        \textwidth 10.in        \topmargin  -.4in
        \headheight 12pt \topskip .4in
        \textheight 6.9in \footskip 0pt
        \def\@oddhead{\thepage\hfil\addtocounter{page}{1}\thepage}
        \let\@evenhead\@oddhead \def\@oddfoot{} \def\@evenfoot{} }

\def\numberbysection{\@addtoreset{equation}{section}
        \def\theequation{\thesection.\arabic{equation}}}

\def\underline#1{\relax\ifmmode\@@underline#1\else
        $\@@underline{\hbox{#1}}$\relax\fi}

\def\titlepage{\@restonecolfalse\if@twocolumn\@restonecoltrue\onecolumn
     \else \newpage \fi \thispagestyle{empty}\c@page\z@
        \def\thefootnote{\fnsymbol{footnote}} }

\def\endtitlepage{\if@restonecol\twocolumn \else \newpage \fi
        \def\thefootnote{\arabic{footnote}}
        \setcounter{footnote}{0}}  

\catcode`@=12
\relax

\def\figcap{\section*{Figure Captions\markboth
        {FIGURECAPTIONS}{FIGURECAPTIONS}}\list
        {Figure \arabic{enumi}:\hfill}{\settowidth\labelwidth{Figure
999:}
        \leftmargin\labelwidth
        \advance\leftmargin\labelsep\usecounter{enumi}}}
 \relax
\def\tablecap{\section*{Table Captions\markboth
        {TABLECAPTIONS}{TABLECAPTIONS}}\list
        {Table \arabic{enumi}:\hfill}{\settowidth\labelwidth{Table
999:}
        \leftmargin\labelwidth
        \advance\leftmargin\labelsep\usecounter{enumi}}}
 \relax
\def\reflist{\section*{References\markboth
        {REFLIST}{REFLIST}}\list
        {[\arabic{enumi}]\hfill}{\settowidth\labelwidth{[999]}
        \leftmargin\labelwidth
        \advance\leftmargin\labelsep\usecounter{enumi}}}
 \relax
\makeatletter
\newcounter{pubctr}
\def\publist{\@ifnextchar[{\@publist}{\@@publist}}
\def\@publist[#1]{\list
        {[\arabic{pubctr}]\hfill}{\settowidth\labelwidth{[999]}
        \leftmargin\labelwidth
        \advance\leftmargin\labelsep
        \@nmbrlisttrue\def\@listctr{pubctr}
        \setcounter{pubctr}{#1}\addtocounter{pubctr}{-1}}}
\def\@@publist{\list
        {[\arabic{pubctr}]\hfill}{\settowidth\labelwidth{[999]}
        \leftmargin\labelwidth
        \advance\leftmargin\labelsep
        \@nmbrlisttrue\def\@listctr{pubctr}}}
 \relax
\makeatother
\newskip\humongous \humongous=0pt plus 1000pt minus 1000pt

\newif\ifdtup

\relax

\def\be{\begin{equation}}
\def\ee{\end{equation}}
\def\ba{\begin{eqnarray}}
\def\ea{\end{eqnarray}}

\def\del{\partial}

\def\r{\rho}
\def\a{\alpha}

\def\b{\beta}

\def\G{\Gamma}

\def\Th{\Theta}
\def\m{\mu}
\def\n{\nu}

\def\Om{\Omega}

\def\L{\Lambda}
\def\s{\sigma} 
\def\S{\Sigma}

\def\cN{{\cal N}}

\def\no{\noindent}

\def\qq{\qquad}

\def\IR{\relax{\rm I\kern-.18em R}}

\def \ha {{1\over 2}}

\def \ov {\over}

\def\IR{\relax{\rm I\kern-.18em R}}
\def\inv{^{\raise.15ex\hbox{${\scriptscriptstyle -}$}\kern-.05em 1}}

\def\tL{{\tilde L}}

\begin{document}

\renewcommand{\theequation}{\arabic{equation}}

\newcommand{\beq}{\begin{equation}}
\newcommand{\eeq}[1]{\label{#1}\end{equation}}
\newcommand{\ber}{\begin{eqnarray}}
\newcommand{\eer}[1]{\label{#1}\end{eqnarray}}
\newcommand{\eqn}[1]{(\ref{#1})}
\begin{titlepage}
\begin{center}

\hfill CERN-TH/99-37\\
\hfill hep--th/9902125\\

\vskip .8in

{\large \bf On running couplings in gauge theories from\\
type-IIB supergravity }

\vskip 0.6in

{\bf A. Kehagias}\phantom{x} and\phantom{x} {\bf K. Sfetsos}
\vskip 0.1in
{\em Theory Division, CERN\\
     CH-1211 Geneva 23, Switzerland\\
{\tt kehagias,sfetsos@mail.cern.ch}}\\
\vskip .2in

\end{center}

\vskip .6in

\centerline{\bf Abstract }

\no
We construct an explicit solution of type-IIB supergravity 
describing the strong coupling regime of a non-supersymmetric gauge theory.
The latter has a running coupling with an ultraviolet stable
fixed point 
corresponding to the $\cN= 4$ $SU(N)$
super-Yang--Mills theory at large $N$. 
The running coupling has a power law behaviour,
argued to be universal, that is consistent 
with holography.
Around the critical point, our solution defines
an asymptotic expansion for the gauge coupling beta-function. 
We also calculate the first correction to the Coulombic quark--antiquark 
potential.

\vskip 0,2cm
\no

\vskip 4cm
\noindent
CERN-TH/99-37\\
February 1999\\
\end{titlepage}
\vfill
\eject

\def\baselinestretch{1.2}
\baselineskip 16 pt
\noindent

\def\tT{{\tilde T}}
\def\tg{{\tilde g}}
\def\tL{{\tilde L}}


\section{Introduction and computations}

One of the well-known vacua of the type-IIB supergravity 
theory is the $AdS_5\times S^5$ one, first described in 
\cite{S}. The non-vanishing fields are the metric and the anti-self-dual 
five-form $F_5$. The latter is given by the Freund--Rubin-type ansatz,
which is explicitly written as
\ba
F_{\mu\nu\rho\kappa\lambda}&=& -\frac{\sqrt{\Lambda}}{2}
 \epsilon_{\mu\nu\rho\kappa\lambda}\ , \qq
\mu,\nu,\ldots=0,1,\dots,4\ , 
\nonumber\\
F_{ijkpq}&=& \frac{\sqrt{\Lambda}}{2} \epsilon_{ijkpq}\ , \qq
i,j,\dots=5,\dots,9\ ,
\label{FF}
\ea
and is clearly anti-self-dual. This background has received a lot of 
attention recently because of its conjectured connection to 
${\cal N}=4$
$SU(N)$ super-Yang--Mills (SYM) theory at large $N$ \cite{M,K}. 
The SYM coupling $g_{\rm YM}$ is given, in terms of the 
dilaton $\Phi$, as  $g^2_{\rm YM}=4\pi e^\Phi$, and  the 't Hooft coupling  is
$g^2_{\rm H}=g^2_{\rm YM} N$, where $N= \int_{S^5}F_5$ is the flux of 
the five-form through the $S^5$.
The dilaton is  constant 
in this background, which is related to the finiteness of the 
${\cal N}=4$ SYM theory. 

In order to make contact with QCD, it is important to investigate 
deformations of the SYM
theory that break conformal invariance and supersymmetry. 
In this case, the couplings are 
running corresponding to a non-constant dilaton in the supergravity 
side. It is then clear that the background we are after is 
a perturbation of $AdS_5\times S^5$. Attempts to 
find supergravity backgrounds that allow a non-constant 
dilaton, and hence a running coupling of the YM theory, have been 
exploited within type-$0$ theories \cite{KT}--\cite{E}. Deformations of the 
${\cal N}=4$ theory, which flow to an interacting conformal fixed point,
have been considered in \cite{DZ}. 

The purpose of the present paper is to show that running couplings are 
also possible within  the type-IIB string theory.
We will study  the ``minimal'' case, that  is we will 
 keep the same $F_5$ 
form as in \eqn{FF} and turn on a non-constant dilaton. We will show 
that  such a solution, which 
 breaks supersymmetry and  conformal invariance, exists.
We will assume for the metric four-dimensional 
Poincar\'e invariance $ISO(1,3)$, 
since we would like a gauge theory defined on Minkowski space-time. 
In addition,
we will preserve the original $SO(6)$ symmetry of the $AdS_5\times S^5$. 
As a result, the $ISO(1,3)\times SO(6)$ invariant ten-dimensional metric is 
of the form\footnote{Supersymmetric solutions to type-IIB supergravity 
which, however,
do not preserve the Poincar\'e invariance in the brane world-volume 
have been found in \cite{ced}.}
$$ds^2=g_{\mu\nu}dx^\mu dx^\nu+  g_{ij}dx^idx^j\ , $$ 
where    
\be 
g_{\mu\nu}dx^\mu dx^\nu = \Om^2(\r) (d\r^2 + dx_\a dx^\a)\ , \qq \a=0,1,2,3\ ,
\label{anz1}
\ee
and $g_{ij}$ is the metric on $S^5$. The dilaton, by 
$ISO(1,3)\times SO(6)$ invariance, can only be a function of $\rho$. 
The supergravity equations turn out to be 
\ba
&&R_{\m\n} = - \L g_{\m\n} + \ha \del_\m \Phi\del_\n \Phi\ ,
\nonumber \\
&& {1\ov \sqrt{-G} } \del_\m \left(\sqrt{-G} G^{\m\n} \del_\n \Phi\right)
 =0 \ ,
\label{eqs1}
\ea
and 
\be
R_{ij} \ =\  \L g_{ij} \ . \label{eqs12}
\ee
The above equation is automatically solved for a five-sphere of radius 
$2/\sqrt{\L}$ and  a first integral of the dilaton equation  in \eqn{eqs1} is
\ba
\del_\r \Phi = A \Om^{-3}\ ,
\label{eqs2}
\ea 
where $A$ is a dimensionful integration constant. 
Moreover, the non-zero components
of the Ricci tensor for the metric \eqn{anz1}  are
\ba
R_{\r\r} & = & - 4 \del_\r^2 \ln \Om \ ,
\nonumber\\
R_{\a\b} & = & - \eta_{\a\b} \left( \del_\r^2 \ln \Om 
+ 3 \big(\del_\r \ln \Om\big)^2 \right)\ ,
\label{ricc1}
\ea
and the first equation in \eqn{eqs1} reduces to solving
\ba
&&\big(\del_\r \ln\Om\big)^2 = {A^2\ov 24}\ \Om^{-6} + {\L\ov 4}\ \Om^2\ .
\label{eqs3}
\ea
The solution of the above equation 
for $\Om$ as a function of $\r$ is given implicitly, 
in terms of a hypergeometric function, by 
\be
\Om^3\ F\Big(\ha,{3\ov 8},{11\ov 8};-{6\L \ov A^2}\Om^8\Big) 
= \pm \sqrt{3\ov 8} |A| ( \r-\r_0) \ , 
\label{sol1}
\ee
where $\r_0$ is another constant of integration.
The different overall signs in the right-hand side of \eqn{sol1} arise 
from taking the square root in \eqn{eqs3}.
We impose the boundary condition that the space described by \eqn{anz1}
becomes $AdS_5$ when $\r\to 0^+$. That means that the conformal factor 
should assume the form $\Om\simeq R/\r$ for small $\r$, where $R\equiv
(4\pi g_s N)^{1/4}$. Using well-known 
formulae for the hypergeometric functions, we see that this naturally 
leads to the 
choice of the minus sign in \eqn{sol1} and also fixes $\L= 4/R^2$.
In addition, the constants $A$ and $\r_0$ are related by
\be
|A| = R^3 {\eta^4\ov \r_0^4}\ ,\qq \eta \equiv
 \sqrt{3\ov 8 \pi} {\G(3/8)\G(1/8)\ov 24^{3/8}}\simeq 1.87\ .
\label{roo}
\ee
Then \eqn{sol1} can be written as
\be
\Big({\Om\r_0\ov R}\Big)^3 F\left(\ha,{3\ov 8},{11\ov 8};
-{24\ov \eta^8} \Big({\Om\r_0\ov R}\Big)^8\right) 
= \sqrt{3\ov 8} \eta^4 \left(1- {\r\ov \r_0}\right) \ . 
\label{sol11}
\ee
We also find that
\be
\Om \simeq (3/8)^{1/6} \eta^{4/3} {R\ov \r_0} 
\left(1-{\r\ov \r_0}\right)^{1/3} \ ,\qq {\rm as} \quad
\r\to \r_0^-\ .
\label{assy1}
\ee
We may solve the dilaton equation in \eqn{eqs2} close to $\r=0$ and
$\r=\r_0$. The result for the string coupling is
\be
e^\Phi = g_s\left(1+s {\eta^4\ov 4} \Big({\r\ov \r_0}\Big)^4 
+ {\eta^8\ov 32} \Big({\r\ov \r_0}\Big)^8 +
s {11 \eta^{12}\ov 3456 } \Big({\r\ov \r_0}\Big)^{12} +
  \dots \right)\ ,\qq
{\rm as} \quad  \r\to 0^+\ ,\label{eph0}
\ee
where $s=\pm 1={\rm sign} (A)$,\footnote{The solutions corresponding to 
the two possible choices for $s$ are related by an $S$-duality 
transformation and correspond to different gauge theories. This reflects
the fact that, except for $\r_0=\infty$, corresponding to the 
$\cN=4$ $SU(N)$ SYM at large $N$, 
our solution describes gauge theories that are not
$S$-duality-invariant. Also in \eqn{eph0} we have used \eqn{asss} below.}
while on the other side
\be
\Phi = \Phi_0 - s\ (8/3)^{1/2} \ln\left(1-\frac{\r}{\r_0}\right)\ ,\qq
{\rm as} \quad  \r\to \r_0^-\ .\label{eph01}
\ee
The form of the above solution  is dictated by the fact that, 
in the limit $\rho_0\to \infty$, the dilaton should be  $e^\Phi=g_s\equiv
e^{\Phi_0}$ and 
\eqn{eph0} and \eqn{eph01} should coincide. 
Now, at $\r=\r_0$ there is a singularity that may 
be easily seen by computing
the Ricci scalar using \eqn{eqs1}. The latter is ${\cal R}=\ha \del_\m \Phi
\del^\m \Phi$, which at $\r\simeq \r_0$ behaves as ${\cal R}
\sim (\r_0-\r)^{-8/3}$.
Hence, we may consider our solution only as an asymptotic expansion around
the $AdS_5$ geometry at $\r=0$ of the form 
\be
\Om(\r) = {R\ov \r}\left(1+
\sum_{n=1}^\infty a_n \Big({\r\ov \r_0}\Big)^{8 n}\right) \ ,
\qq \r< \r_0  \ ,
\label{asss}
\ee
where the coefficients $a_n$ are computed using the series representation
for the hypergeometric function in \eqn{sol11} for large $\Om$. 
The first coefficient of the expansion turns out to be 
$a_1=-\eta^8/432\simeq -0.352$. Hence, $\Om(\r)$ is given by
\be
\Om(\r) \simeq {R\ov \r} \left( 1- {\eta^8\ov 432} \Big({\r\ov \r_0}\Big)^8
\right)
\label{assa}
\ee
to a very good approximation, except when $\r$ takes values very close 
to $\r_0$. Then we may use, to a very good approximation as well, \eqn{assy1}
instead of \eqn{assa}, and the results are plotted in fig. 1. 
Note that our analysis was done in the Einstein frame and it 
is not difficult to translate everything into the string frame
by multiplying the Einstein metric by $e^{\Phi/2}$. Our solution
is singular at $\rho=\rho_0$ in both frames and can, indeed, be trusted 
away from that point.  

Note that, the metric \eqn{anz1} with the conformal factor $\Om$ 
specified by \eqn{sol11}
can be written in horospherical coordinates\footnote{We thank A.A. Tseytlin
for comments on this point. Also, a similar solution to that 
in the present paper, but with no analogous interpretation, 
has been given in \cite{NO}.} $(r,x^\a)$ determined by 
$\Om(\r) d\r = dr$, as
\be
g_{\m\n} dx^\m dx^\n = dr^2 + K(r) dx_\a dx^\a\ ,
\label{horr1}
\ee
where 
\be 
K(r) = \sqrt{2} R^2 e^{-2 r_0/R} \sinh^{1/2}\left(4(r_0-r)\ov R\right)\ .
\label{hor2}
\ee
In the same coordinate system the string coupling takes the form
\be
e^\Phi =g_s \left( \coth \Big({2(r_0-r)\ov R}\Big)\right)^{2s a}\ .
\qq a\equiv  \sqrt{6}/4\ .
\label{hor3}
\ee
In the rest of the paper we prefer to work in the $(\r,x^\a)$ coordinate
system.
\begin{figure}[ht]
\bigskip
\centerline{ \epsfxsize 4.0 truein \epsfbox {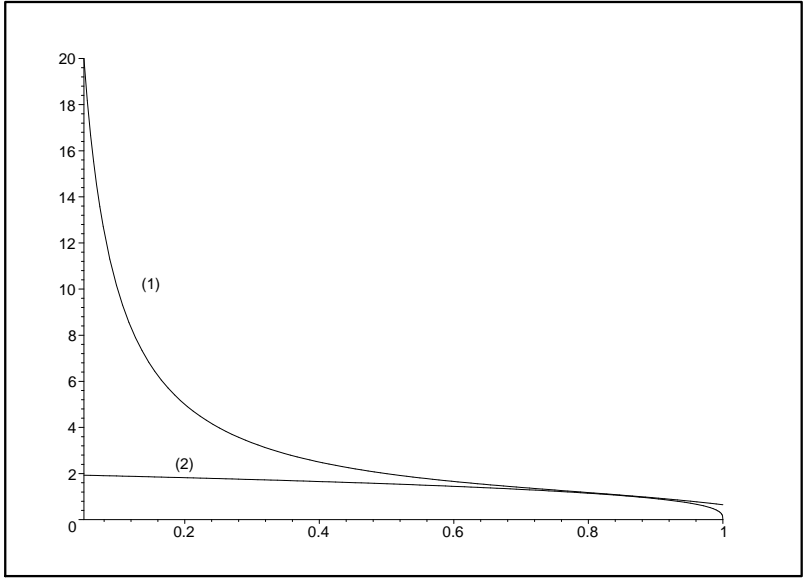}}
\caption{Plot of $\Om(\r)/R$ in units where 
$\rho_0=1$. Curves $(1)$ and $(2)$ were plotted using \protect\eqn{assa} and
\protect\eqn{assy1} respectively. The curve corresponding to $\Om(\r)/R$, 
obtained by numerically solving \protect\eqn{sol1}, coincides 
with the union of these curves.}
\end{figure}

\section{Running Coupling}

In the AdS/CFT scheme,  the dependence of the bulk fields on the
radial coordinate $\rho$ may be interpreted as energy dependence. In fact, 
it is a general feature in the AdS/CFT scheme that long (short) distances in 
the AdS space correspond to high (low) energies in the CFT
\cite{M,SW}. In particular,
if the dilaton in the supergravity side is a function of $\rho$,
then the 't Hooft coupling of the boundary CFT has an  energy 
dependence
and can be interpreted as the running coupling of the CFT. Running coupling 
means of course that we are away from conformality; thus, backgrounds that 
admit non-constant dilaton correspond to non-conformal field theories.
As long as supersymmetry is unbroken,  
spin-sum rules for the AdS supersymmetry are expected to protect 
the 't Hooft coupling $g_{\rm H}$ 
of the boundary ${\cal N}=4$ YM theory against 
running. However, if supersymmetry is broken, 
there are then no more cancellations between fermionic and bosonic 
contributions leading to the running of  $g_{\rm H}$. 
The  specific background we found here clearly breaks supersymmetry,
and  
\begin{eqnarray}
\delta\lambda&=&\frac{1}{2}\gamma^M\partial_M\Phi\epsilon^*=
\frac{A}{2\Omega^3}\gamma^\rho\epsilon^*\ ,
\nonumber\\
\delta \psi_M&=&D_M\epsilon
\end{eqnarray}
are the associated fermionic zero modes.  
If we now follow the correspondence between long-distances/high-energies 
in the AdS/CFT scheme, we find that the dual theory of the supergravity 
solution we obtained has a coupling with power-law running. Indeed, 
by changing the variable $\r=R^2/U$ and interpreting $U$ as the energy of the 
boundary field theory, we find from \eqn{eph0}  the running of $g_H$:
\be
g_{\rm H}=g_{\rm H}^{*}
\left(1+s \frac{\eta^4}{8}\frac{R^8}{\rho_0^4 U^4}
+ \frac{\eta^8}{128}\frac{R^{16}}{\rho_0^8 U^8} + s {17 \eta^{12}\ov 27648}
{R^{24}\ov \r_0^{12} U^{12}} + \dots \right)\ , 
\label{runn}
\ee
where $g_{\rm H}^*=R^2$ is the UV value of the 't Hooft coupling (the 
result in plotted in fig. 2).
%
%
\begin{figure}[ht]
\epsfxsize=4in
\centerline{\epsffile{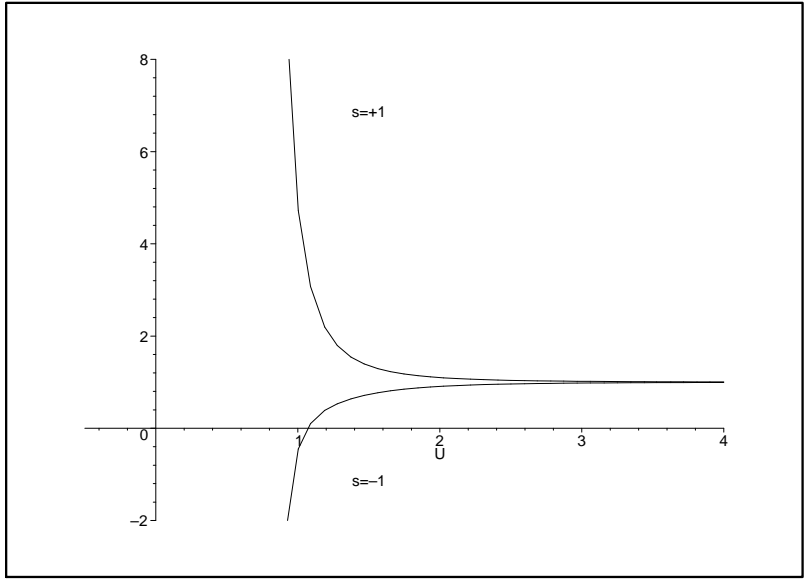}}
\caption{Plot of $g_{\rm H}/g^*_{\rm H} $ as a function of $U$ using 
\eqn{runn}.}
\end{figure}
From this expression, it may easily be found that the behaviour 
of the beta-function for the 't Hooft coupling  around $g_{\rm H}^*$ 
is 
\be 
U {d g_{\rm H}\ov dU} = -4 (g_{\rm H} - g^*_{\rm H})
- 2  (g_{\rm H} - g^*_{\rm H})^2/g^*_{\rm H}
-{14\ov 27} (g_{\rm H} - g^*_{\rm H})^3/g^{*2}_{\rm H}
\ +\
{\cal O} (g_{\rm H} - g^*_{\rm H})^4 \ .
\label{e1tta}
\ee
However, the above
equation does not specify the beta-function, but rather its derivative at the 
$g_{\rm H}= g^*_{\rm H}$ point. The reason is that our solution breaks down 
at energies $U\sim R^2/\r_0$. From \eqn{e1tta} we see that
\be 
\b'(g^*_{\rm H}) =-4\ , \label{beta}
\ee
which means that $g^*_{\rm YM}$ is a UV-stable fixed point.\footnote{
Using a radious/energy relation in horospherical coordinates 
of the form $U= R^2 e^{-r/R}$ we find that the running of $g_{\rm H}$ is 
\be
U {d g_{\rm H}\ov dU}= - ag^*_{\rm H}
 \left( \Big( { g_{\rm H}\ov g^*_{\rm H}}\Big)^{1+
{1\ov a}}  -\Big( { g_{\rm H}\ov g^*_{\rm H}}\Big)^{1-
{1\ov a} }\right) \ .
\label{e1ta2}
\ee 
However, the above expression is trustable only around the fixed point 
$ g^*_{\rm H}$.
}
We believe
that \eqn{beta} is universal, namely, that it is  
valid for all models that approach $AdS_5 \times S^5$ at some 
boundary. This can be seen by recalling that, near $AdS_5 \times S^5$,
the dilaton always satisfies \eqn{eqs1} with $\Omega=R/\rho$. As a result,
$\Phi$ will behave as $e^{\Phi-\Phi_0}\sim \rho^4$, where $\Phi_0$ is the 
value of $\Phi$ at $\rho=0$.  We also see 
that \eqn{e1tta}  determines the second and third 
derivatives of the coupling
beta-function at the fixed point, which, however, are not expected 
to be  model-independent. Let us also note that there is no known perturbative
field theory with UV-stable fixed points. 
A behaviour of the form \eqn{runn}, 
namely power-law running of the couplings,
was also found in type-0 theories (see second ref. in \cite{E}),
in gauge theories in higher dimensions \cite{TV}
and extensively 
discussed in  gauge-coupling unification in 
theories with large internal dimensions \cite{ant}--\cite{B}. 
In this scenario, the internal 
dimensions are shown up in the four-dimensional theory as 
the massive KK modes. These  modes can run in the loops of the 
four-dimensional theory, giving rise to a power law running of the couplings. 
In particular, for $d$ large extra dimensions and for energies $E$ above the 
infrared cutoff, which is specified by the mass scale $\mu$ of 
the extra dimensions, 
we find, just for dimensional reasons, that the running 
coupling constant of the effective four-dimensional theory is of the form 
$g^{(4)}\!-\!g_0^{(4)}\sim (\mu/E )^{d/2}$, 
where $g_0^{(4)}$ is the bare coupling. 
Thus, in our case, since we have a 
four-dimensional theory coming from ten dimensions, we should expect 
the coupling to run in the sixth power of $E$. Instead, we find here
that the coupling depends on the fourth power of $U=E$, indicating that when 
holography is involved, we get a softer running of the couplings.

It is possible to identify those operators that are 
responsible for the running 
of the coupling in the boundary field theory. 
Since   the dilaton approaches 
a constant value at $\rho=0$ and the asymptotic background is an  $AdS_5$ 
space, the corresponding  
boundary field theory is expected to be a deformation of 
the ${\cal N}=4$ $SU(N)$ 
supersymmetric YM theory. The explicit form of the deformation may be
specified by recalling that our solution still has an $SO(6)$ symmetry.
There are not many $SO(6)$ singlets in the spectrum of the $S^5$ 
compactification. In fact, from the results in \cite{GM} we see 
that the only scalar singlets are  
the complex scalar of type-IIB theory $B$,
$a_{ijkl}$ and $h^i_i$, with masses $m^2_B=0,\, m^2_a=m^2_h=32$ in AdS-mass
units. Since we have not perturbed the five-form $F_5$, $a_{ijkl}=0$ and thus 
the perturbations we have turned on are the real part of $B$ and $h_i^i$. 
From their masses we find that the former corresponds 
to marginal deformations of the type $F_{\mu\nu}^2$, while the latter
corresponds, to the 
dimension-eight operator $F_{\mu\nu}^4$, which is irrelevant. 
However, it gives 
contributions to the boundary field theory since we have an IR cutoff 
specified  by $\rho_0$. Sending $\rho_0\to \infty$ all bulk perturbations 
disappear and, similarly, the boundary field theory turns out to be the 
${\cal N}=4$ large-$N$ $SU(N)$ SYM theory.

\section{The quark--antiquark potential}

The breaking of the superconformal invariance of the $\cN=4$ theory by
our solution should be apparent in 
the expression for the quark--antiquark potential, which we now compute
along the lines of \cite{Mal1,Rey}. 
We will find corrections to the purely Coulombic behaviour, which, on purely 
dimensional grounds, we expect to be in powers of $A L^4$, 
where $L$ is the quark--antiquark distance.
We are eventually interested in the first such correction, 
which, as is apparent when comparing \eqn{asss} with \eqn{eph0},
is due to the dilaton, but for the moment we keep the formalism general.
As usual, we have to minimize the Nambu--Goto action 
\be
S={1\ov 2\pi} \int d\tau d\s\ \sqrt{\det\left(G_{MN}\del_\a X^M
\del_\b X^N\right)}
\ ,
\label{NaG}
\ee
where $G_{MN}$ is the target-space metric in the string frame.
For the static configuration $x^0=\tau$, $x^1\equiv x=\s$, and $x^2,x^3$
as well as the coordinates of $S^5$ held fixed, we find that \eqn{NaG}
becomes (we use the notation of \cite{Mal1})
\be
S = {T\ov 2\pi} \int dx\ e^{\Th/2} \sqrt{(\del_x U)^2 +{U^4/R^4}} \ ,
\label{acts}
\ee
where $e^\Th\equiv e^\Phi \S^4$ and 
$\S=1+\sum_{n=1}^\infty a_n \Big({R^2\ov \r_0 U}\big)^{8n}$ 
is the function multiplying $R/\r$ in \eqn{asss} (rewritten using $\r=R^2/U$).
It is clear that any background that 
approaches $AdS_5$ will always have a $\S$ of the form
\eqn{asss}, so that our analysis is quite general at this point.
It is easy to see that the solution is expressed as 
\be
x=  \int_{U_0}^U {R^2 dU\ov U^2 \sqrt{e^{\Th-\Th_0} U^4/U_0^4-1}}\ ,
\label{xlu}
\ee
where $U_0$ is the smallest ``distance'' of the trajectory to the center 
and $\Th_0$ is the value of the function $\Th$ evaluated at $U_0$.
We assume that one of the branes is taken out to $U=\infty$ and that the 
string configuration starts and ends at this brane. The rest of the branes 
are located at $U=R^2/\r_0$. 
Setting $x=L/2$ corresponds to $U=\infty$. In turn, this
gives a condition that relates $U_0$ to $L$ as
\be
L= 2  \int_{U_0}^\infty {R^2 dU\ov U^2 \sqrt{e^{\Th-\Th_0} U^4/U_0^4-1}}\ .
\label{luo}
\ee
Proceeding in a standard fashion, we substitute back our solution \eqn{xlu}
into the action \eqn{acts} and obtain an integral that is infinity.
This is because we have included into the potential energy the 
(equal) masses of the infinitely heavy (in the supergravity approximation)
quark and antiquark. In order to 
compute these masses, we assume that $N$-branes are at $U=R^2/\r_0$ and 1
at $U=U_{\rm max}$ that is assumed to be large but finite.
The mass of a single quark is computed if in \eqn{acts} we consider a 
configuration with $x^0=\tau$, $U=\s$ and with fixed spatial world-volume 
coordinates $x^\a$, $\a=1,2,3$, as well as $S^5$ coordinates.
Then the self-energy of the quark is
\be 
E_{\rm self}= {1\ov 2\pi} \int_{R^2/\r_0}^{U_{\rm max}} dU 
e^{\Th/ 2}\ .
\label{jdhh}
\ee
Subtracting off this energy twice and letting $U_{\rm max}\to \infty$, 
we obtain a finite result for the
quark--antiquark potential given by 
\be
E_{q\bar q}= {1\ov \pi} \int^{\infty}_{U_0} dU 
{U^2/U_0^4e^{\Th/2} \ov \sqrt{U^4/U_0^2-e^{\Th_0-\Th}}}  - {1\ov \pi}
\int_{R^2/\r_0}^{\infty} dU e^{\Th/ 2}\ .
\label{eeee}
\ee
At this point we have to solve \eqn{luo} for $U_0$ as a function of $L$
and substitute the result back into \eqn{eeee} to obtain $E_{q\bar q}$ as
a function of $L$ only. This can be done perturbatively in powers of $L/\r_0$,
and we are interested in the first correction to the Coulombic law
behaviour of the potential. As explained, the correction due to 
the non-constant dilaton is dominant and we will therefore use $\Om=R/\r$, 
which corresponds to $AdS_5\times S^5$ for the string metric,
whereas for the string coupling we will keep 
the first two terms in \eqn{eph0}. 
Then we find that 
\be
U_0\simeq {2R^2 \eta_1\ov L} 
\left( 1+ {s\ov 8} \Big({\eta\ov 2 \eta_1}\Big)^2 \Big({L\ov \r_0}\Big)^4
\right)\ ,\qq \eta_1 ={\pi^{1/2} \Gamma(3/4)\ov \G(1/4)}\simeq 0.599\ ,
\label{luo1}
\ee
where the value of the numerical constant $\eta$ has been given in \eqn{roo}
and
\be
E_{\rm self} \simeq {U_{\rm max}\ov 2\pi} -{R^2\ov 2\pi \r_0} 
\left(1- s {\eta^4\ov 24}\right)\ .
\label{hdhh1}
\ee
The result for the quark--antiquark potential is
\be
E_{q\bar q}\simeq  - {2 \eta_1^2 R^2\ov \pi} {1\ov L}\left(1- {s\ov 8} 
\Big({\eta\ov 2 \eta_1}\Big)^4 \Big({L\ov \r_0}\Big)^4 \right)
+ {R^2\ov \pi \r_0}\left(1-s {\eta^4\ov 24}\right) \ . 
\label{eeee1}
\ee
We see that the Coulombic potential receives a correction proportional to
$L^3$ due to the breaking of conformal invariance.\footnote{
Supergravity is valid when $\r\ll \r_0$, which means that 
$U\geq U_0\gg R^2/\r_0$. Using the leading term in \eqn{luo},
we deduce that $L/\r_0\ll 1$, which is indeed the condition for the validity
of \eqn{eeee1}.}
This is remarkably
similar to the potential obtained in \cite{BISY1} for the quark--antiquark 
pair for $\cN=4$,
at finite temperature, using the near-horizon supergravity solution for
$N$ coincident $D3$-branes.  
However, in that case supersymmetry is broken by 
thermal effects, whereas in our case it is broken, 
by the presence of a non-trivial dilaton, even at zero temperature.
The last term in \eqn{eeee1} does not depend on $L$ and represents a constant
shift of the potential energy.

As a final remark we note that the computation of the potential for 
the monopole--antimonopole pair proceeds along the same lines as that for
the quark--antiquark pair,
with the only difference that we start, similarly to \cite{GO}, 
with the action for a D-string.
This means that the integrand in \eqn{acts} should be multiplied by
$e^{-\Phi}$. Hence, the function $e^\Th$ entering into \eqn{acts} is defined 
as $e^\Th=e^{-\Phi}\S^4$. Consequently,  
the first correction to the Coulombic behaviour of the 
monopole--antimonopole potential is given by
\be
E_{m\bar m}\simeq 
-{2 \eta_1^2 R^2\ov\pi g_s L}\left(1+  {s\ov 8} 
\Big({\eta\ov 2 \eta_1}\Big)^4 \Big({L\ov \r_0}\Big)^4 \right)
+ {R^2\ov \pi \r_0}\left(1+s {\eta^4\ov 24}\right) \ , 
\label{hghh}
\ee
which is the same as \eqn{eeee1} after we use the fact that 
under S-duality $g_s\to 1/g_s$ and $s\to -s$. 
Hence we see a screening
(antiscreening) of the quark--antiquark pair for $s=+1$($-1$) and
exactly the opposite behaviour for the monopole--antimonopole pair.
 
\bigskip\bigskip
\centerline{ \bf Acknowledgements}

We would like to thank  K. Dienes, E. Dudas, T. Gherghetta, A.A. Tseytlin
and A. Zaffaroni for discussions.




\begin{thebibliography}{99}
\bibitem{S} J.H. Schwarz, Nucl. Phys. {\bf B226} (1983) 269.

\bibitem{M}J. Maldacena, Adv. Theor. Math. Phys. {\bf 2} (1998) 231, 
hep-th/9711200.

\bibitem{K} S.S. Gubser, I.R. Klebanov and  A.M. Polyakov,
 Phys. Lett. {\bf B428} (1998)105, hep-th/9802109;
E. Witten, Adv. Theor. Math. Phys. {\bf 2} (1998) 253,
hep-th/9802150.

\bibitem{KT} I.R. Klebanov and A.A. Tseytlin, {\it 
D-branes and dual gauge theories in type 0 strings}, hep-th/9811035, 
{\it Asymptotic freedom and infrared behaviour in the type 0
string approach to gauge theories},  hep-th/9812089;
{\it A non-supersymmetric large-$N$ CFT from type 0 string theory}, 
hep-th/9901101.
\bibitem{Mh}  J.A. Minahan, {\it Glueball mass spectra and other issues for supergravity duals of QCD models}, hep-th/9811156; {\it Asymptotic freedom 
and confinement from type 0 string theory}, hep-th/9902074.


\bibitem{E} 
G. Ferretti and D. Martelli, {\it On the construction of gauge theories 
from non-critical type 0 strings},  hep-th/9811208; 
E. Alv\'arez and C. G\'omez, {\it Non-critical confining 
strings and the renormalization group}, hep-th/9902012.  



\bibitem{DZ} J. Distler and  F. Zamora, 
{\it  Non-supersymmetric conformal field theories from stable Anti-de sitter 
spaces}, hep-th/9810206; L. Girardello, 
M. Petrini, M. Porrati and A. Zaffaroni,
{ \it Novel local CFT and exact results on perturbations of N=4 
super Yang-Mills from AdS dynamics}, hep-th/9810126;
A. Karch, D. L\"ust and 
A. Miemiec, {\it New $N=1$ superconformal field theories
and their supergravity description}, hep-th/9901041.

\bibitem{ced}
M. Cederwall, U. Gran, M. Holm and B.E.W. Nilsson,
{\it Finite tensor deformations of supergravity solitons}, hep-th/9812144.

\bibitem{NO}
S. Nojiri and S.D. Odintsov, Phys. Lett. {\bf B449} (1999) 39, hep-th/9812017.

\bibitem{SW} L. Susskind and E. Witten, {\it The holographic bound in Anti-de 
Sitter space}, hep-th/9805114;
A.W. Peet and J. Polchinski, {\it UV/IR 
relations in AdS dynamics}, hep-th/9809022.   

\bibitem{TV}
T.R. Taylor and G. Veneziano, Phys. Lett. {\bf B212} (1988) 147.

\bibitem{ant}
I. Antoniadis, Phys. Lett. {\bf B246} (1990) 377.

\bibitem{WL}
E. Witten, Nucl. Phys. {\bf B471} (1996) 135, hep-th/9602070; J.D. Lykken, 
Phys. Rev. {\bf D54} (1996) 3693, hep-th/9603133. 
 

\bibitem{DDG} K.R. Dienes, E. Dudas and T. Gherghetta, Phys. Lett. 
{\bf B436} (1998) 55, hep-ph/9803466; Nucl. Phys. {\bf B537} (1999) 47, 
hep-ph/9806292. 
\bibitem{B} C. Bachas, JHEP {\bf 23} (1998) 9811, hep-ph/9807415.


\bibitem{GM} M G\"unaydin and N. Marcus, 
Class. Quant. Grav. {\bf 2} (1985) L11; H.J. Kim, L.J. Romans and 
P. van Nieuwenhuizen, Phys. Rev. {\bf D32} (1985) 389.



\bibitem{Mal1}
J. Maldacena,
Phys. Rev. Lett. {\bf 80} (1998) 4859, hep-th/9803002. 

\bibitem{Rey}
S-J. Rey and J. Yee, {\it Macroscopic strings as heavy quarks in large N
gauge theory and Anti-de-Sitter supergravity}, hep-th/9803001. 


\bibitem{BISY1}
A. Brandhuber, N. Itzhaki, J. Sonnenschein and S. Yankielowicz,
Phys. Lett. {\bf B434} (1998) 36, hep-th/9803137. 
 

\bibitem{GO}
D.J. Gross and H. Ooguri, 
Phys. Rev. {\bf D58} (1998) 106002, hep-th/9805129. 

\end{thebibliography}
\end{document}